\documentclass[a4paper,11pt]{article}
\usepackage{pos}
\usepackage{multirow}

\title{Resolving the NO$\nu$A and T2K tension in the presence of Neutrino Non-Standard interactions}
\author*[a]{Sabya Sachi Chatterjee}
\author[b,c]{Antonio Palazzo}

\affiliation[a]{Institut de Physique Th\'{e}orique, Universit\'{e} Paris Saclay, CNRS, CEA, F-91191 Gif-sur-Yvette, France}

\affiliation[b]{Dipartimento Interateneo di Fisica``Michelangelo Merlin," Via Amendola 173, 70126 Bari, Italy}
\affiliation[c]{Istituto Nazionale di Fisica Nucleare, Sezione di Bari, Via Orabona 4, 70126 Bari, Italy}

\emailAdd{sabya-sachi.chatterjee@ipht.fr}
\emailAdd{palazzo@ba.infn.it}

\abstract{The current data of the two long-baseline accelerator experiments NO$\nu$A and T2K, shows a tension at more than 90$\%$ C.L. for 2 degrees of freedom, in the determination of the standard CP-phase $\delta_{\mathrm {CP}}$ in case of neutrino normal ordering (NO). NO$\nu$A measures the value close to $\delta_{\mathrm {CP}} \sim 0.8 \pi$, while T2K prefers the value of $\delta_{\mathrm {CP}} \sim 1.4 \pi$. We show that such a tension can be resolved if one hypothesizes the existence of neutral-current non-standard interactions (NSI) of neutrinos involving the flavor changing type $e-\mu$ or the $e-\tau$ sectors with couplings $|\varepsilon_{e\mu}| \sim |\varepsilon_{e\tau}|\sim 0.2$. Remarkably, our analyses show that in the presence of such NSI, both the experiments point towards the same common value of the standard CP-phase $\delta_{\mathrm {CP}} \sim 3\pi/2$, thereby indicating towards the maximal CP-violation in the standard $3\nu$ framework. We also show that the best fit values of the new CP-phases $\phi_{e\mu}$ or $ \phi_{e\tau}$ are close to $\sim 3\pi/2$, hence pointing towards the maximal CP-violation in the NSI sector.}

\FullConference{%
  *** The 22nd International Workshop on Neutrinos from Accelerators (NuFact2021) ***\\
  *** 6–11 Sep 2021 ***\\
  *** Cagliari, Italy ***}

\begin{document}
\maketitle

{\bf {\em Introduction.}}  The new datasets\cite{NOVA_talk_nu2020,T2K_talk_nu2020} from the two long-baseline (LBL) accelerator experiments NO$\nu$A and T2K were recently presented at the Neutrino 2020 Conference. Interestingly, the measurements on the standard 3-flavor CP-phase $\delta_{\rm CP}$ show an appreciable discrepancy between the two experiments in case of NO, and this lies at more than $90\%$ C.L. for 2 degrees of freedom (d.o.f.). 
This discrepancy can be attributed either to a statistical fluctuation or to an unknown systematic error,or it may be the signature of some new physics beyond the Standard Model (SM). In particular, these two experiments are of different nature with respect to their sensitivity to the matter effects due to their different baselines (810 km for NO$\nu$A and 295 km for T2K). This invites ample of opportunities for the new physics to play an exciting role. In this work we explore the impact of neutral current (NC) non-standard interactions (NSI) of neutrinos in resolving the tension between the two experiments.

{\bf {\em Theoretical framework.}} NSI represents the low-energy manifestation of high-energy physics involving the new
heavy states (for a review see~\cite{Farzan:2017xzy,Dev:2019anc}) or, 
the light mediators~\cite{Farzan:2015hkd}.
As first recognised in~\cite{Wolfenstein:1977ue}, NSI of type neutral current (NC) can alter the dynamics of the neutrino flavor oscillation in matter and  
%The presence of NSI can have a sizeable impact on the interpretation of current LBL data. Notably, in the recent work~\cite{Capozzi:2019iqn}, it has been evidenced that they may even obscure the correct determination of the neutrino mass ordering (NMO).%
%\footnote{In the 3-flavor scheme there are three mass eigenstates $\nu_i$ with masses
%$m_i\, (i = 1,2,3)$, three mixing angles $\theta_{12},\theta_{13}, \theta_{13}$, and one CP-phase $\delta_{\mathrm {CP}}$.
%The mass ordering is defined to be normal (inverted) if $m_3>m_{1,2}$  ($m_3<m_{1,2}$).
%We will abbreviate normal (inverted) ordering as NO (IO).}
%The impact of NSI on present and future new-generation LBL experiments has been widely explored (see for example~\cite{Friedland:2012tq,Coelho:2012bp,Girardi:2014kca,Rahman:2015vqa,Coloma:2015kiu,deGouvea:2015ndi,Agarwalla:2016fkh,Liao:2016hsa,Forero:2016cmb,Huitu:2016bmb,Bakhti:2016prn,Masud:2016bvp,Soumya:2016enw,Masud:2016gcl,deGouvea:2016pom,Fukasawa:2016lew,Liao:2016orc,Liao:2016bgf,Blennow:2016etl,Deepthi:2017gxg,Flores:2018kwk,Hyde:2018tqt,Masud:2018pig,Esteban:2019lfo}.)
this can be represented by a dimension-six operator~\cite{Wolfenstein:1977ue}
%%%%%%%%%%%%%%%%%%%%%%%%%%%%%%%%%%%%%%%%%%%%
\begin{equation}
\mathcal{L}_{\mathrm{NC-NSI}} \;=\;
-2\sqrt{2}G_F 
%\sum_{\alpha\beta f}
\varepsilon_{\alpha\beta}^{fC}
\bigl(\overline{\nu_\alpha}\gamma^\mu P_L \nu_\beta\bigr)
\bigl(\overline{f}\gamma_\mu P_C f\bigr)
%+ h.c.
\;,
\label{H_NC-NSI}
\end{equation}
%%%%%%%%%%%%%%%%%%%%%%%%%%%%%%%%%%%%%%%%%%%
where $\alpha, \beta = e,\mu,\tau$ denote the 
neutrino flavor,  $f = e,u,d$ indicate the matter 
fermions, $P$ represents the projector operator with
superscript $C=L, R$ referring to the chirality of the 
$ff$ current, and $\varepsilon_{\alpha\beta}^{fC}$ are the strengths 
of the NSI. The hermiticity of the interaction implies
%%%%%%%%%%%%%%%%%%%%%%%%%%%%%%%%%%%%%%%%%%%%%%%%%%%%%%
%\begin{equation}
$\varepsilon_{\beta\alpha}^{fC} \;=\; (\varepsilon_{\alpha\beta}^{fC})^*$.\,
%\;.
%\end{equation}
%%%%%%%%%%%%%%%%%%%%%%%%%%%%%%%%%%%%%%%%%%%%%%%%%%%%%%
For the neutrino propagation in matter, the effective NSI couplings can be written as
%%%%%%%%%%%%%%%%%%%%%%%%%%%%%%%%%%%%%%%%%%%%%%%%%%%%%%
\begin{equation}
\varepsilon_{\alpha\beta}
\;\equiv\; 
\sum_{f,C}
\varepsilon_{\alpha\beta}^{fC}
\dfrac{N_f}{N_e}
\;\equiv\;
\sum_{f=e,u,d}
\left(
\varepsilon_{\alpha\beta}^{fL}+
\varepsilon_{\alpha\beta}^{fR}
\right)\dfrac{N_f}{N_e}
\;,
\label{epsilondef}
\end{equation}
%%%%%%%%%%%%%%%%%%%%%%%%%%%%%%%%%%%%%%%%%%%%%%%%%%%%%%
$N_f$ being the number density of $f$ fermion.
For the neutral and isoscalar Earth matter, $N_n \simeq N_p = N_e$, 
which in turn leads to $N_u \simeq N_d \simeq 3N_e$. So,
%%%%%%%%%%%%%%%%%%%%%%%%%%%%%%%%%%%%%%%%%%%%%%%%%%%%%%
%\begin{equation}
$\varepsilon_{\alpha\beta}\, \simeq\,
\varepsilon_{\alpha\beta}^{e}
+3\,\varepsilon_{\alpha\beta}^{u}
+3\,\varepsilon_{\alpha\beta}^{d}$
\;.
%\label{epsilon_eff}
%\end{equation}
%%%%%%%%%%%%%%%%%%%%%%%%%%%%%%%%%%%%%%%%%%%%%%%%%%%%%
The effective Hamiltonian which governs the neutrino flavors oscillations through the matter gets modified in the presence of NSI, for more details see~\cite{Chatterjee:2020kkm}.  
In this work we focus only on the flavor changing non-diagonal NSIs $|\varepsilon_{e\mu}|$
and $|\varepsilon_{e\tau}|$ 
%More specifically, we consider the couplings $\varepsilon_{e\mu}$
%and $\varepsilon_{e\tau}$, which, as will we discuss below, introduce a dependency 
%on their associated CP-phase in the appearance $\nu_\mu \to \nu_e$ probability%
%%%%%%%%%%%%%%%%%%%%%%%%%%%%%%%%%%%%%%%%%%%%%%%%%%%%%%%
\footnote{$|\varepsilon_{\mu\tau}|$ is strongly constrained by the atmospheric neutrinos, $|\varepsilon_{\mu\tau}| < 8.0 \times 10^{-3}$~\cite{Aartsen:2017xtt}.} 
along with their associated CP-phases $\phi_{e\mu}$ and $\phi_{e\tau}$ respectively which is a crucial ingredient
to resolve the discrepancy between NO$\nu$A and T2K we are considering.
%%%%%%%%%%%%%%%%%%%%%%%%%%%%%%%%%%%%%%%%%%%%%%%%%%%%%%% 
 Let us focus on the conversion probability relevant for the LBL experiments T2K and NO$\nu$A.
In the presence of NSI, the probability can be  expressed as 
the sum of three terms $P_{\mu e}  \simeq  P_{\rm{0}} + P_{\rm {1}}+   P_{\rm {2}}$
%.........................................................................................................................................
%\begin{eqnarray}
%\label{eq:Pme_4nu_3_terms}
%P_{\mu e}  \simeq  P_{\rm{0}} + P_{\rm {1}}+   P_{\rm {2}}\,,
%\end{eqnarray}
%.........................................................................................................................................
which, using a compact notation take the following forms  
%.........................................................................................................................................
\begin{eqnarray}
\label{eq:P0}
 & P_{\rm {0}} &\,\, \simeq\,  4 s_{13}^2 s^2_{23}  f^2\,,\\
\label{eq:P1}
 & P_{\rm {1}} &\,\,  \simeq\,   8 s_{13} s_{12} c_{12} s_{23} c_{23} \alpha f g \cos({\Delta + \delta_{\mathrm {CP}}})\,,\\
 \label{eq:P2}
 & P_{\rm {2}} &\,\,  \simeq\,  8 s_{13} s_{23} v |\varepsilon|   
 [a f^2 \cos(\delta_{\mathrm {CP}} + \phi) + b f g\cos(\Delta + \delta_{\mathrm {CP}} + \phi)]\,,
\end{eqnarray}
%........................................................................................................................................
where $\Delta \equiv  \Delta m^2_{31}L/4E$ is the atmospheric oscillating frequency,
$L$ is the baseline and $E$ the neutrino energy, $\alpha \equiv \Delta m^2_{21}/ \Delta m^2_{31}$, and $v = 2 V_{CC}E/\Delta m_{31}^2$. Here $V_{CC}=\sqrt{2}G_FN_e$ is the charged current matter potential.
For brevity, we have used the notation ($s_{ij} \equiv \sin \theta_{ij} $, $c_{ij} \equiv \cos \theta_{ij}$), 
and also we have, 
%.........................................................................................................................................
\begin{eqnarray}
\label{eq:S}
& f \equiv &\frac{\sin [(1-v) \Delta]}{1-v}\,, \qquad  g \equiv \frac{\sin v\Delta}{v}\,\\
&a = & s^2_{23}, \quad b = c^2_{23} \quad {\mathrm {if}} \quad \varepsilon = |\varepsilon_{e\mu}|e^{i{\phi_{e\mu}}}\,,\\
&a = & s_{23}c_{23}, \quad b = -s_{23} c_{23} \quad {\mathrm {if}} \quad \varepsilon = |\varepsilon_{e\tau}|e^{i{\phi_{e\tau}}}
\end{eqnarray}
%.........................................................................................................................................
%In Eq.~(\ref{eq:P2}) we have assumed for the NSI coupling the general complex form
%%.........................................................................................................................................
%\begin{eqnarray}
%\varepsilon_{\alpha \beta} = |\varepsilon_{\alpha \beta}|  e^{i\phi_{\alpha \beta}}\,.
%\end{eqnarray}
%%.........................................................................................................................................
%The expression of $P_2$ is different for $\varepsilon_{e\mu}$ and  $\varepsilon_{e\tau}$ and,
%%.........................................................................................................................................
%in Eq. (\ref{eq:P2}), one has to make the replacements
%.........................................................................................................................................
%\begin{eqnarray}
% \label{eq:P2_NSI_1}
% a = s^2_{23}, \quad b = c^2_{23} \quad &{\mathrm {if}}& \quad \varepsilon = |\varepsilon_{e\mu}|e^{i{\phi_{e\mu}}}\,,\\
%  \label{eq:P2_NSI_2}
% a =  s_{23}c_{23}, \quad b = -s_{23} c_{23} \quad &{\mathrm {if}}& \quad \varepsilon = |\varepsilon_{e\tau}|e^{i{\phi_{e\tau}}}\,.
%\end{eqnarray}
% ............................................................................................................................................
In the expressions given in Eqs.~(\ref{eq:P0})-(\ref{eq:P2}) for $P_0$, $P_1$ and $P_2$, 
the sign of $\Delta$, $\alpha$ and $v$ is positive (negative) for normal (inverted) ordering. Similarly for antineutrinos, the sign of all the CP-phases and of the matter parameter $v$ are flipped. 
Finally, we observe that the third term $P_{\rm {2}}$ purely arises because of the 
(complex) NSI coupling and it is different from zero only in matter (i.e. if $v \ne 0$). 
It represents the interference between the matter potential
 $\varepsilon_{e\mu}V_{CC}$  (or $\varepsilon_{e\tau}V_{CC}$)
 with the atmospheric wavenumber $\Delta m_{31}^2/2E$. % 
%%%%%%%%%%%%%%%%%%%%%%%%%%%%%%%%%%%%%%%%%%%%%%%%%
%\footnote{Interestingly, an analogous splitting $P_{\mu e}  \simeq  P_{\rm{0}} + P_{\rm {1}}+   P_{\rm {2}}$
%of the transition probability is valid in the presence of 
%oscillations driven by a sterile neutrino~\cite{Klop:2014ima}. In that case, however, the term $P_2$ 
%emerges due to the interference between the amplitude driven by the
%atmospheric mass difference and that by the mass difference
%corresponding to the sterile neutrino, instead of the interference
%with the term originated from the matter potential.}.
%%%%%%%%%%%%%%%%%%%%%%%%%%%%%%%%%%%%%%%%%%%%%%%%% 

%{\bf {\em  Data used in the analysis.}}  We extracted the datasets of NO$\nu$A and T2K from the latest data released 
%  in~\cite{NOVA_talk_nu2020} and~\cite{T2K_talk_nu2020}.
%  We fully incorporate  both the disappearance and appearance channels in both experiments. 
% In our analysis we use the software GLoBES~\cite{Huber:2004ka,Huber:2007ji} 
%and its additional public tool~\cite{Kopp:NSI}, which can implement NSI.
%In our analysis we have marginalized over $\theta_{13}$  with 3.4\% 1 sigma prior with central value 
%$\sin^2\theta_{13}= 0.0219$ as determined by Daya Bay~\cite{Adey:2018zwh}.
% We have fixed the solar parameters $\Delta m^2_{21}$ and $\theta_{12}$
%at their best fit values estimated in the recent global analysis~\cite{Capozzi:2017ipn}.

\begin{figure}[t!]
\vspace*{-0.0cm}
\hspace*{-0.0cm}
\includegraphics[height=4.cm,width=3.7cm]{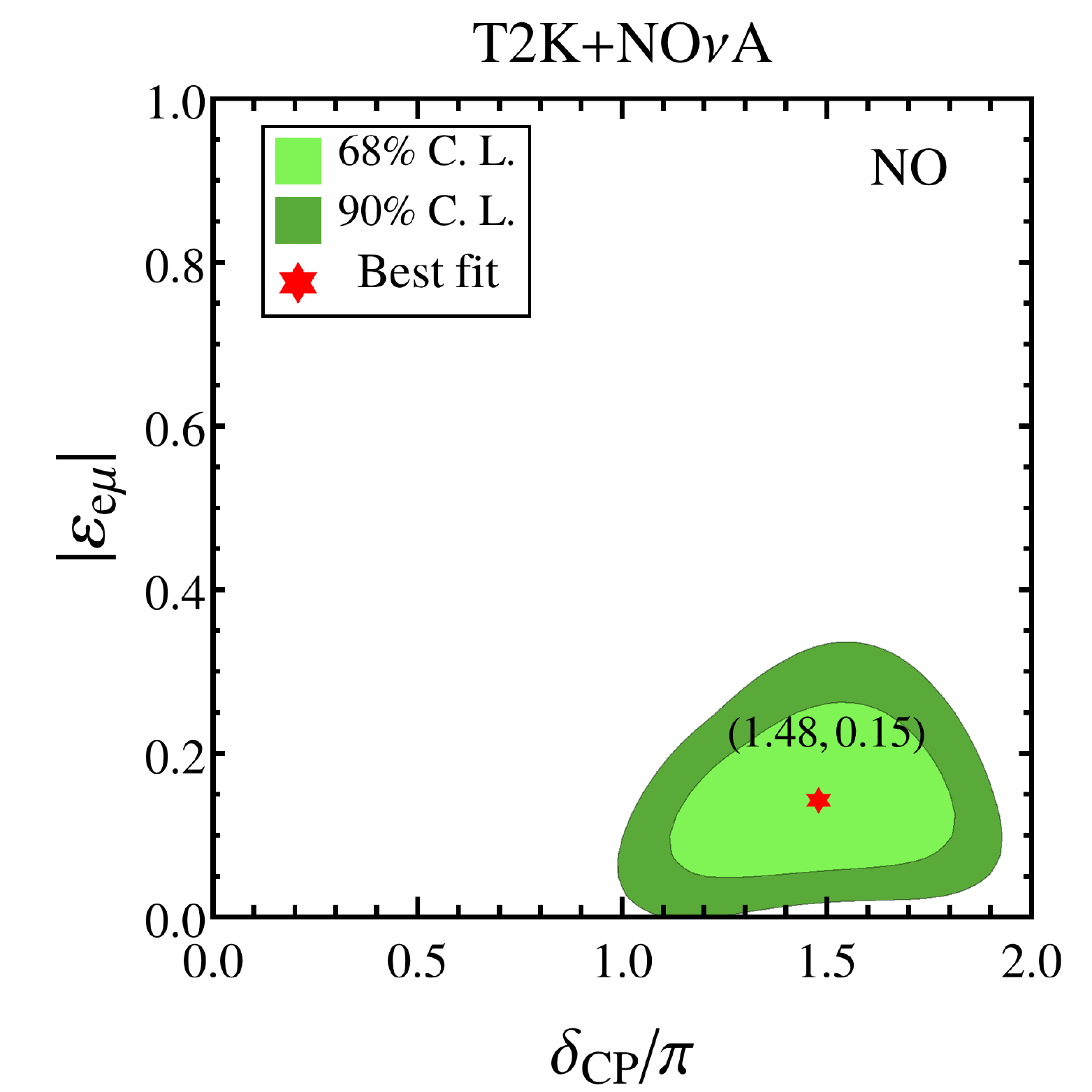}
\includegraphics[height=4.cm,width=3.7cm]{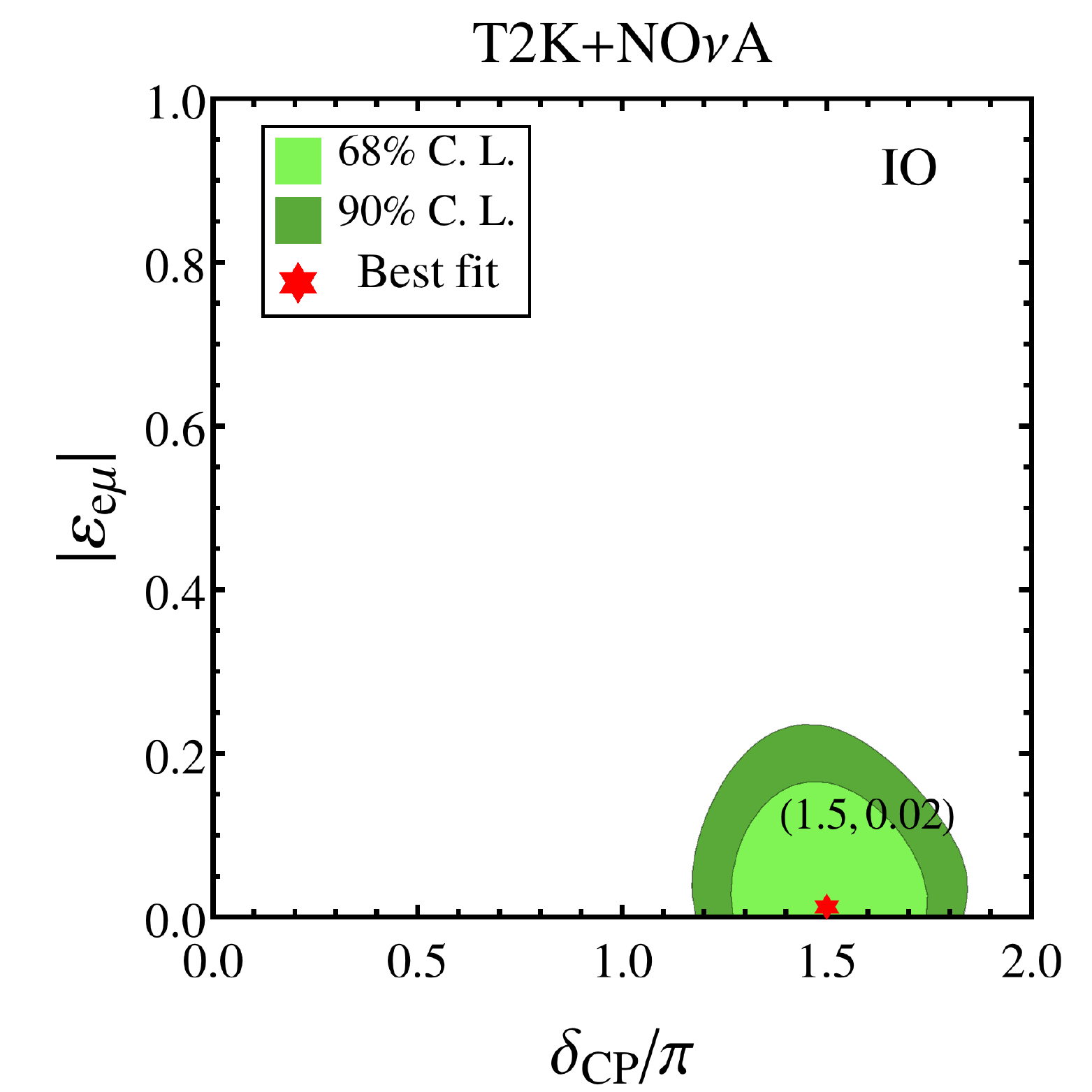}
\includegraphics[height=4.cm,width=3.7cm]{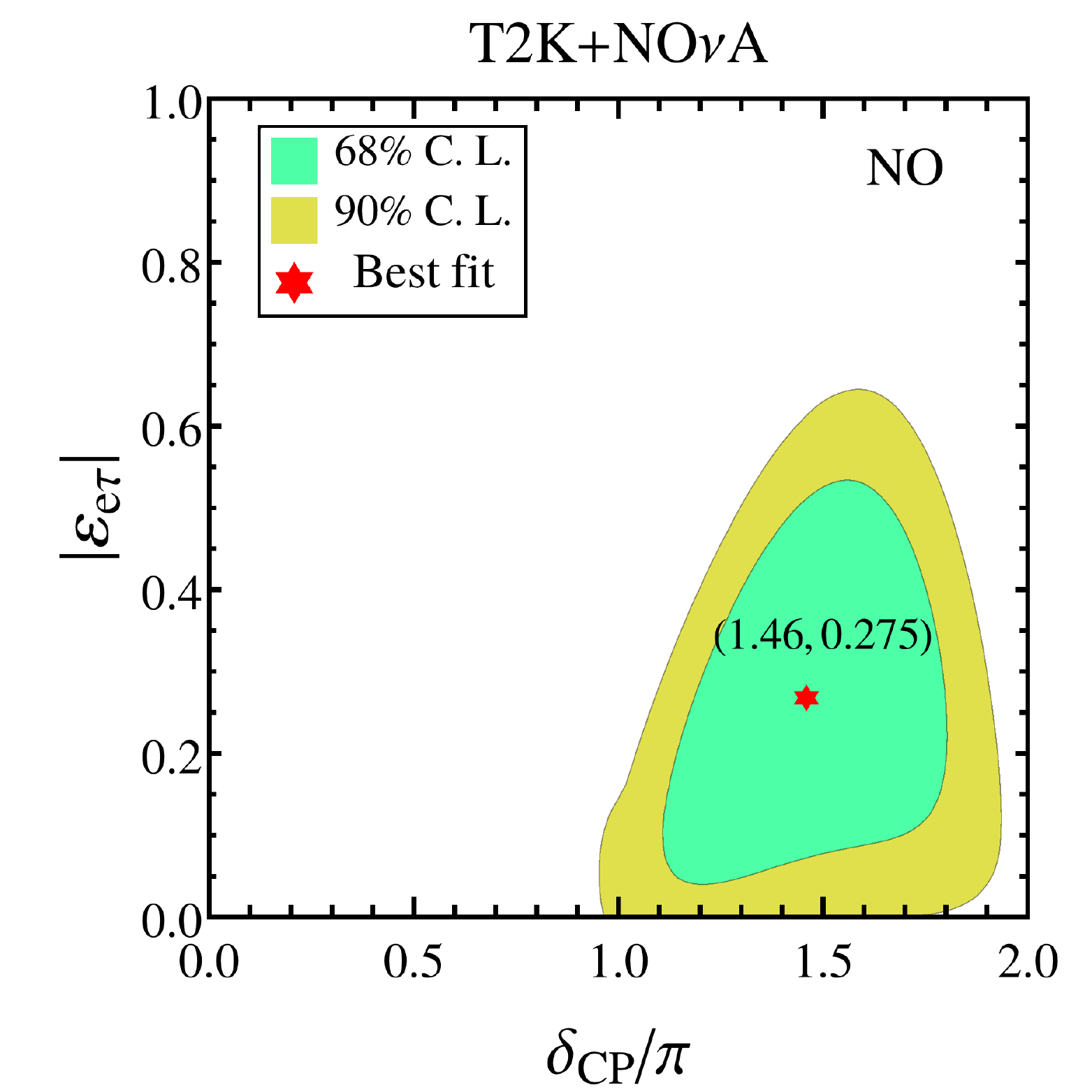}
\includegraphics[height=4.cm,width=3.7cm]{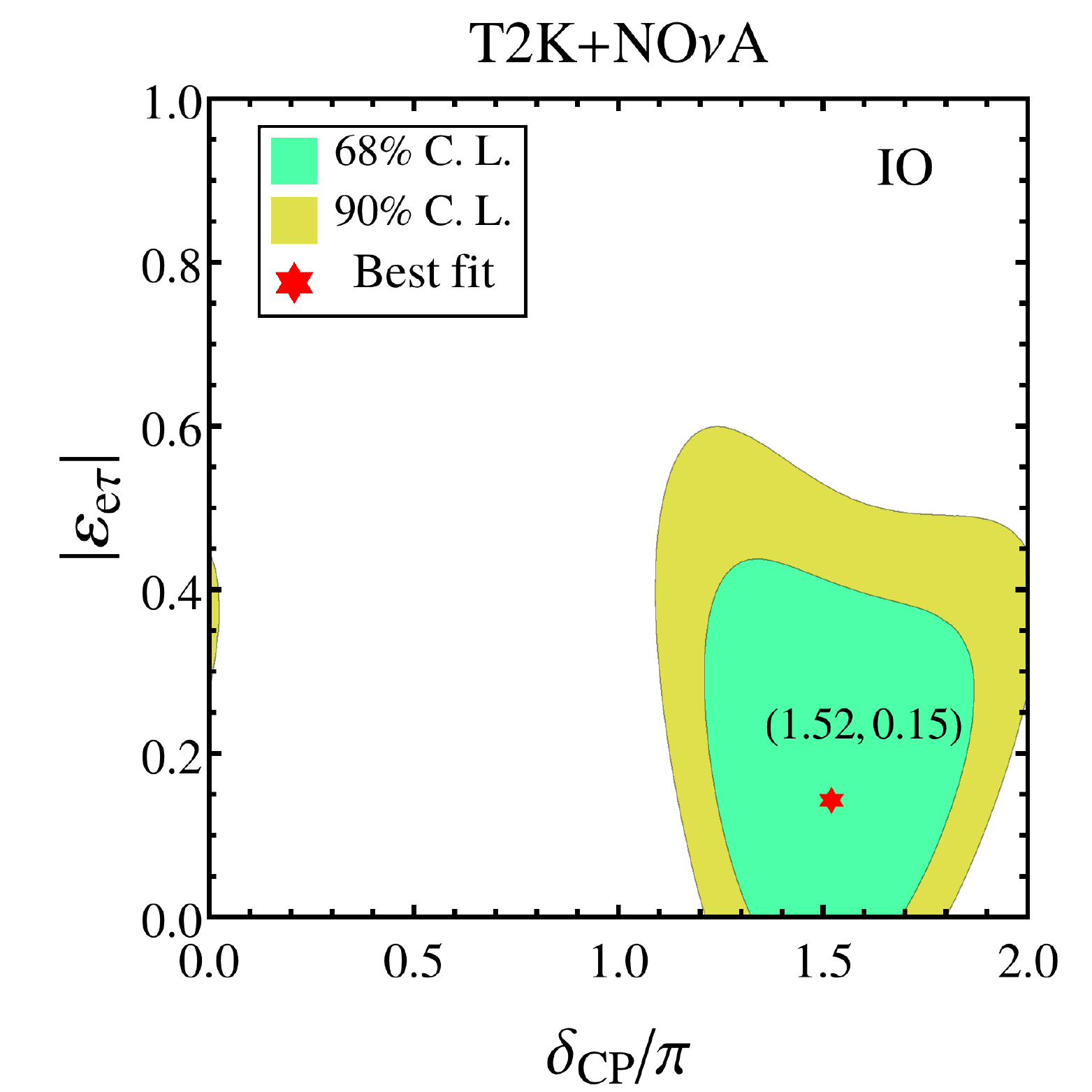}
%\vspace*{-0.3cm}
\caption{Allowed regions obtained from the combined analysis of T2K and NO$\nu$A in the plane $\left(\delta_{\rm CP}/\pi,\,|\varepsilon_{e\mu}|\right)$ and $\left(\delta_{\rm CP}/\pi,\,|\varepsilon_{e\tau}|\right)$ for both the NO and IO. The contours have been drawn at the 68\% and 90\% confidence level for 2 d.o.f.. This figure has been taken from \cite{Chatterjee:2020kkm}.}
\label{fig:regions_1}
\end{figure}

{\bf {\em Numerical Results.}} Figure~\ref{fig:regions_1} reports the results of the combined analyses 
of T2K and NO$\nu$A for both the NO and inverted ordering (IO) in the plane of $\left[|\varepsilon_{e\mu}|, \delta_{\rm CP} \right]$ and $\left[|\varepsilon_{e\tau}|, \delta_{\rm CP} \right]$ respectively. It is worth to note that NSI parameter $\varepsilon_{e\mu}$ or $\varepsilon_{e\tau}$ has been taken one at a time. The contours in each panel are shown at 68\% and 90\% confidence level for 2 d.o.f., where the bestfit point is denoted by a red star. The non-standard CP-phases, the mixing angles $\theta_{13}$ and $\theta_{23}$, and the squared-mass $\Delta m^2_{31}$ have been marginalized away. 
 From the left most panel, one can see that in case of NO, the non zero value of the NSI coupling $|\varepsilon_{e\mu}|$ is preferred over the standard 3-flavor framework with best fit $|\varepsilon_{e\mu}| =0.15$. The statistical significance lies at $\sim$ 2.1$\sigma$ ($\Delta \chi^2 = 4.50$). However in case of IO, the preference of non zero $|\varepsilon_{e\mu}|$ is negligible. Now in case of $|\varepsilon_{e\tau}|$ and NO (third panel), there is a 1.9$\sigma$ ($\Delta \chi^2 = 3.75$) preference of non zero $|\varepsilon_{e\tau}|$ with best fit $|\varepsilon_{e\tau}| =0.27$, while in case of IO (fourth panel) the preference is only at the 1.0$\sigma$ with best fit $|\varepsilon_{e\tau}| =0.15$. It is interesting to note that in all the panels of Fig.~\ref{fig:regions_1}, the preferred best fit values of $\delta_{\rm CP}$ is close to $3\pi/2$, indicating maximal CP-violation in the standard $3\nu$ sector. For more details about the analyses, see~\cite{Chatterjee:2020kkm}. 

%==================================================================
\begin{figure}[t!]
\vspace*{-0.0cm}
\hspace*{-0.0cm}
\includegraphics[height=4.cm,width=3.7cm]{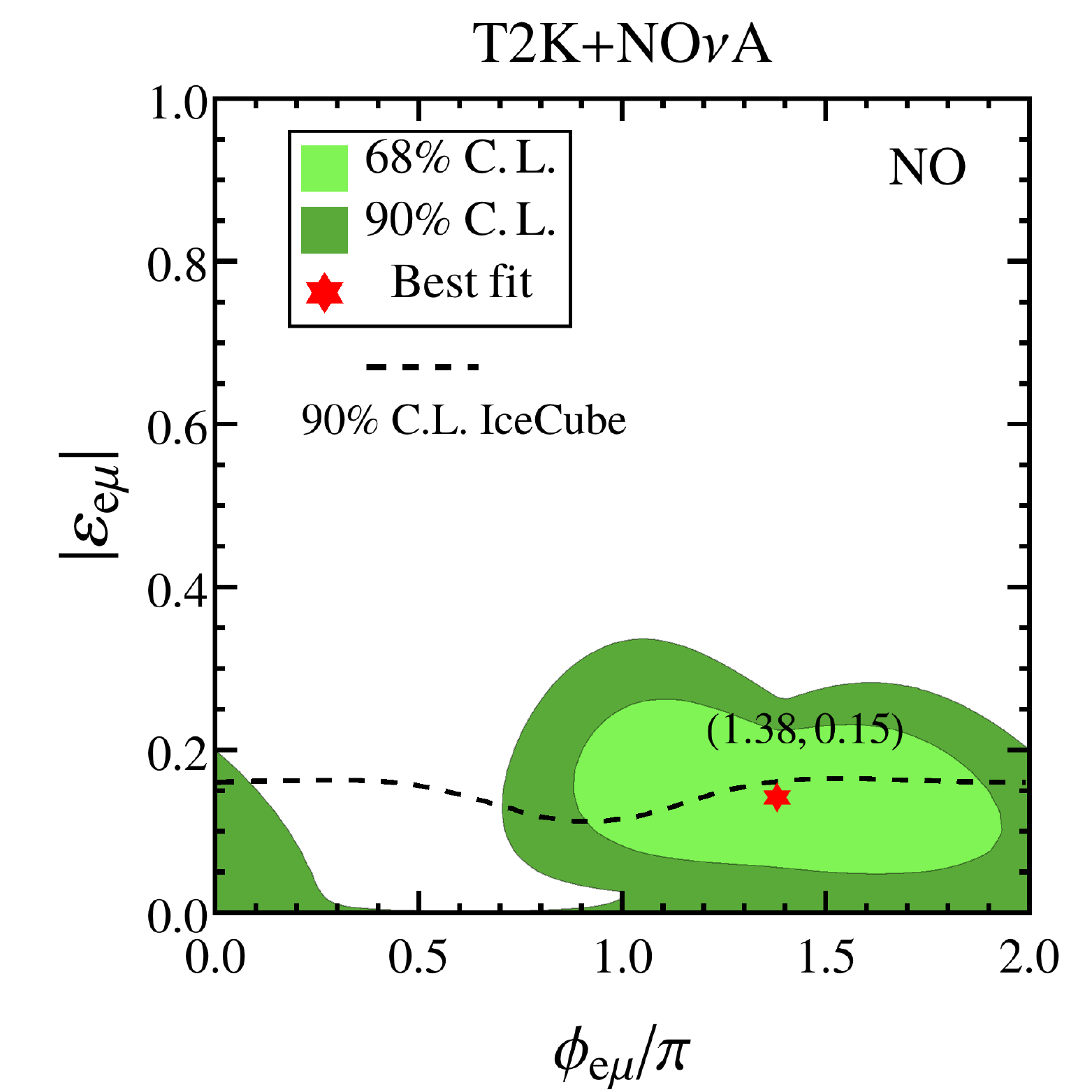}
\includegraphics[height=4.cm,width=3.7cm]{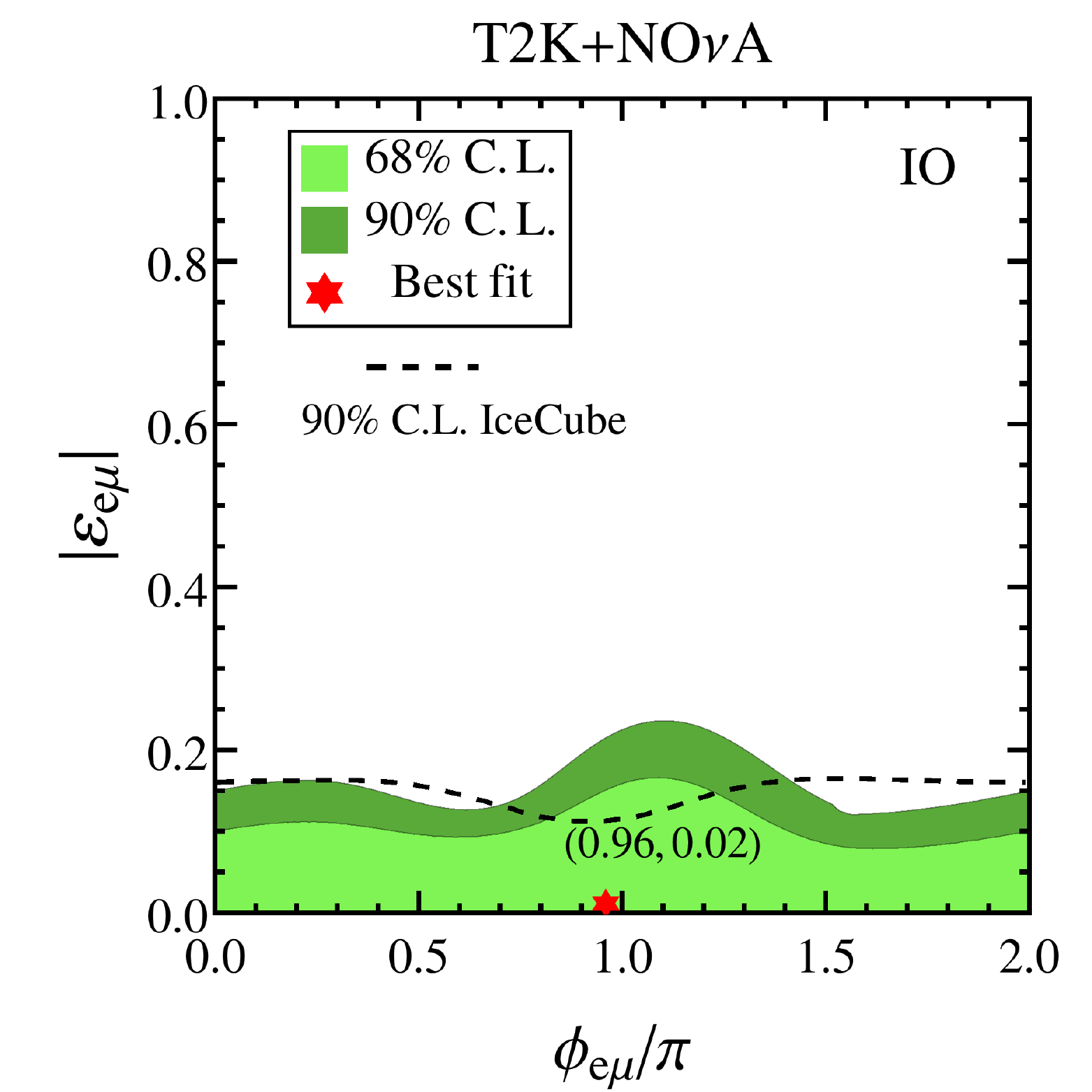}
\includegraphics[height=4.cm,width=3.7cm]{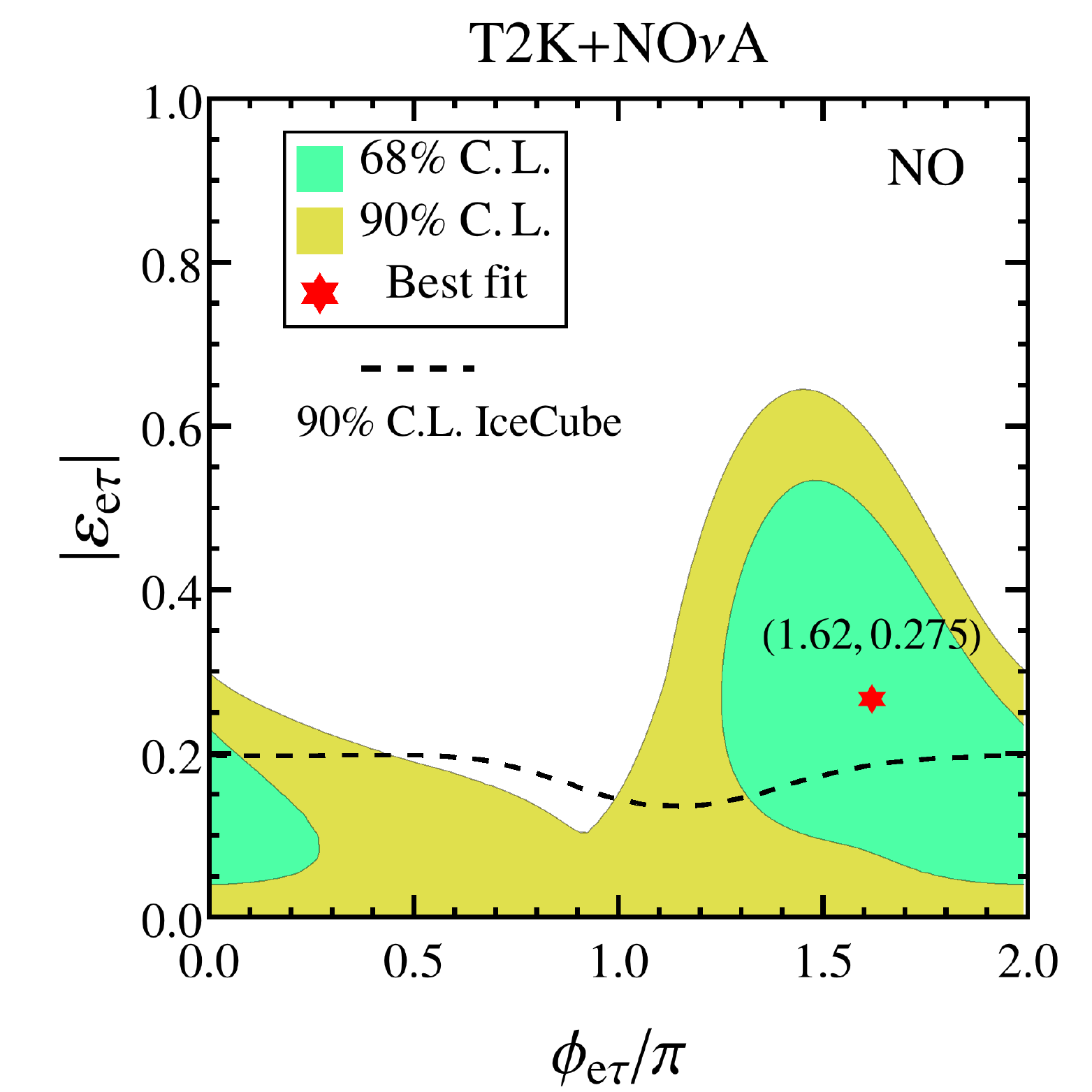}
\includegraphics[height=4.cm,width=3.7cm]{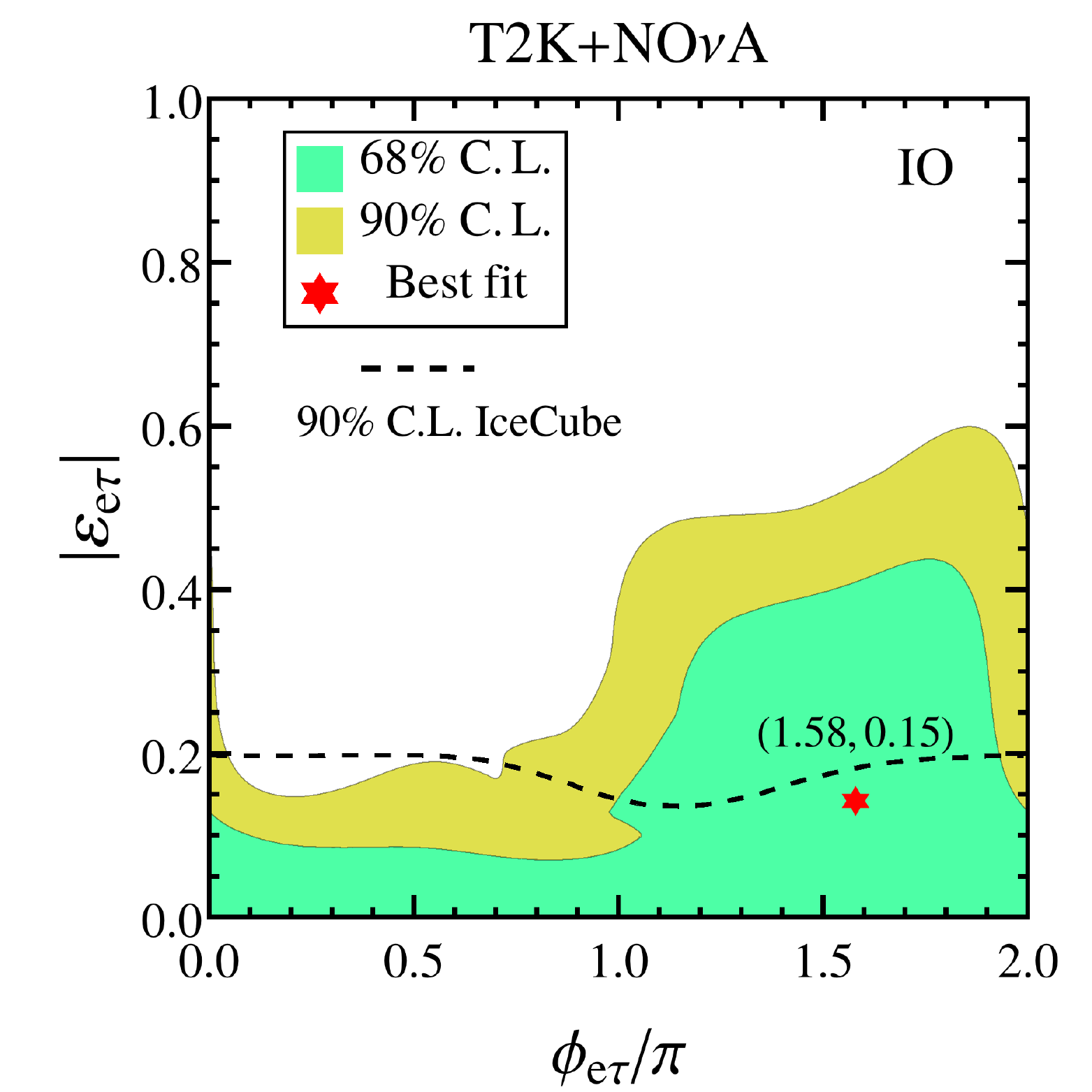}
%\vspace*{-0.3cm}
\caption{Allowed regions determined from the combined analysis of T2K and NO$\nu$A in the plane $\left(\phi_{\rm e\mu}/\pi,\,|\varepsilon_{e\mu}|\right)$ and $\left(\phi_{\rm e\tau}/\pi,\,|\varepsilon_{e\tau}|\right)$ for both the NO and IO. The contours have been drawn at the 68\% and 90\% confidence level for 2 d.o.f.. The dashed curves correspond to the upper bounds (90\% C.L., 2 d.o.f.) derived from the IceCube data~\cite{IceCube_talk_PPNT2020}. This figure has been taken from \cite{Chatterjee:2020kkm}.}
\label{fig:regions_2}
\end{figure}

Figure~\ref{fig:regions_2} represents the contours from the combined analysis of T2K and NO$\nu$A similar to 
Fig.~\ref{fig:regions_1} for both the NO and IO in the plane of $\left[\varepsilon_{e\mu}, \phi_{e\mu} \right]$
and $\left[\varepsilon_{e\tau}, \phi_{e\tau} \right]$ respectively. The standard CP-phase $\delta_{\mathrm {CP}}$,
the mixing angles $\theta_{23}$ and $\theta_{13}$, and the squared-mass $\Delta m^2_{31}$ are marginalized away.
Interestingly in case of NO, the preferred value for both the new CP-phases 
$\phi_{e\mu}$ and $\phi_{e\tau}$ is close to $3\pi/2$, so indicating maximal CP-violation also in the NSI sector. The black dashed line in each panel corresponds to the upper bounds given by the analysis of the IceCube data~\cite{IceCube_talk_PPNT2020}. It can be seen that the bounds we have obtained from the combined analysis, are compatible with that of IceCube. 

%In Table~\ref{table:chi2} we report the best fit values of the NSI couplings
%together with the CP-phases and the value of $\Delta\chi^2=\chi^2_{\rm SM}-\chi^2_{\rm SM + NSI}$ 
%for a fixed choice of the NMO.
%==================================================================

%==================================================================
%==================================================================

%==================================================================
\begin{figure*}[t!]
\vspace*{-0.0cm}
\hspace*{-0.1cm}
\includegraphics[height=5.87cm,width=5.cm]{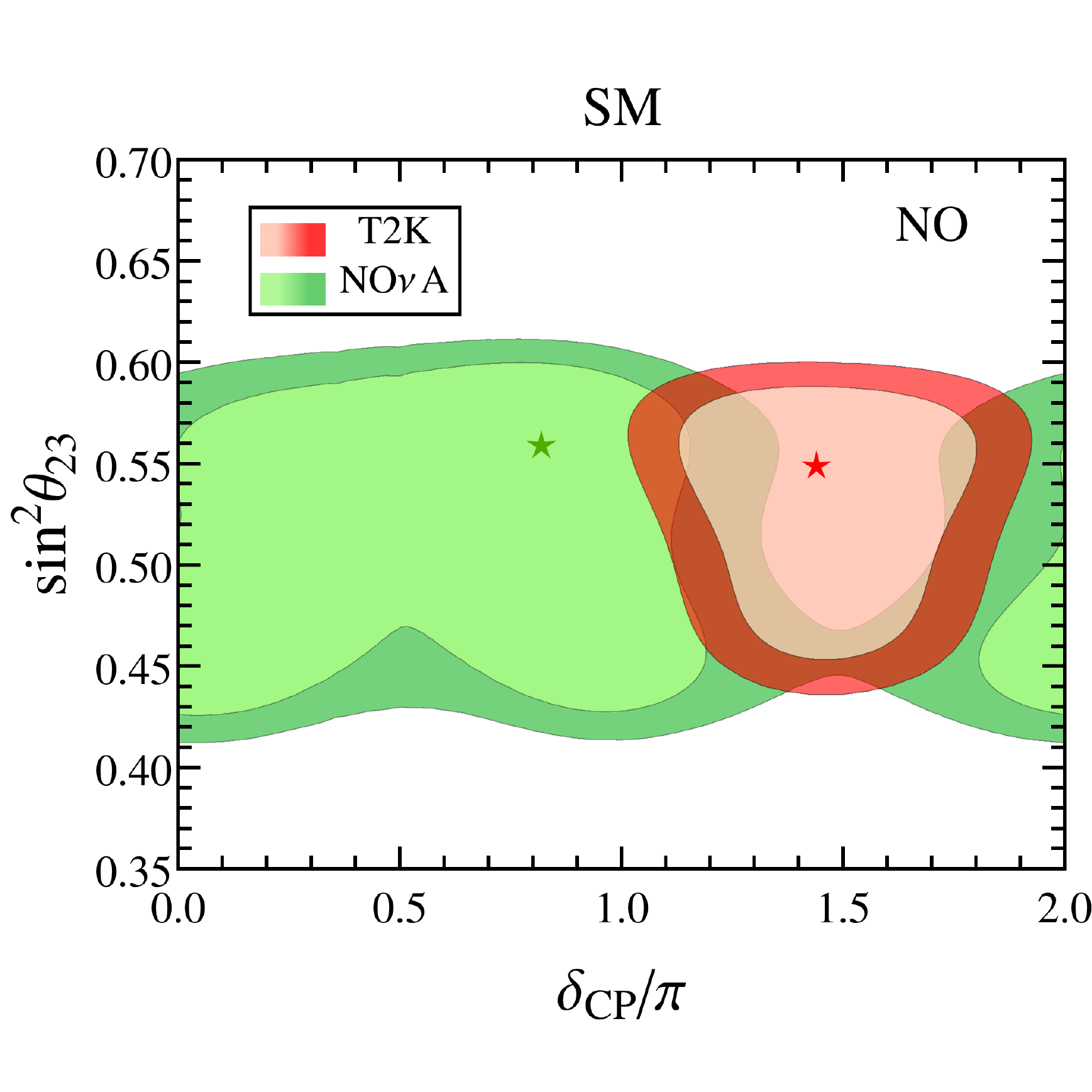}
\includegraphics[height=5.87cm,width=5.cm]{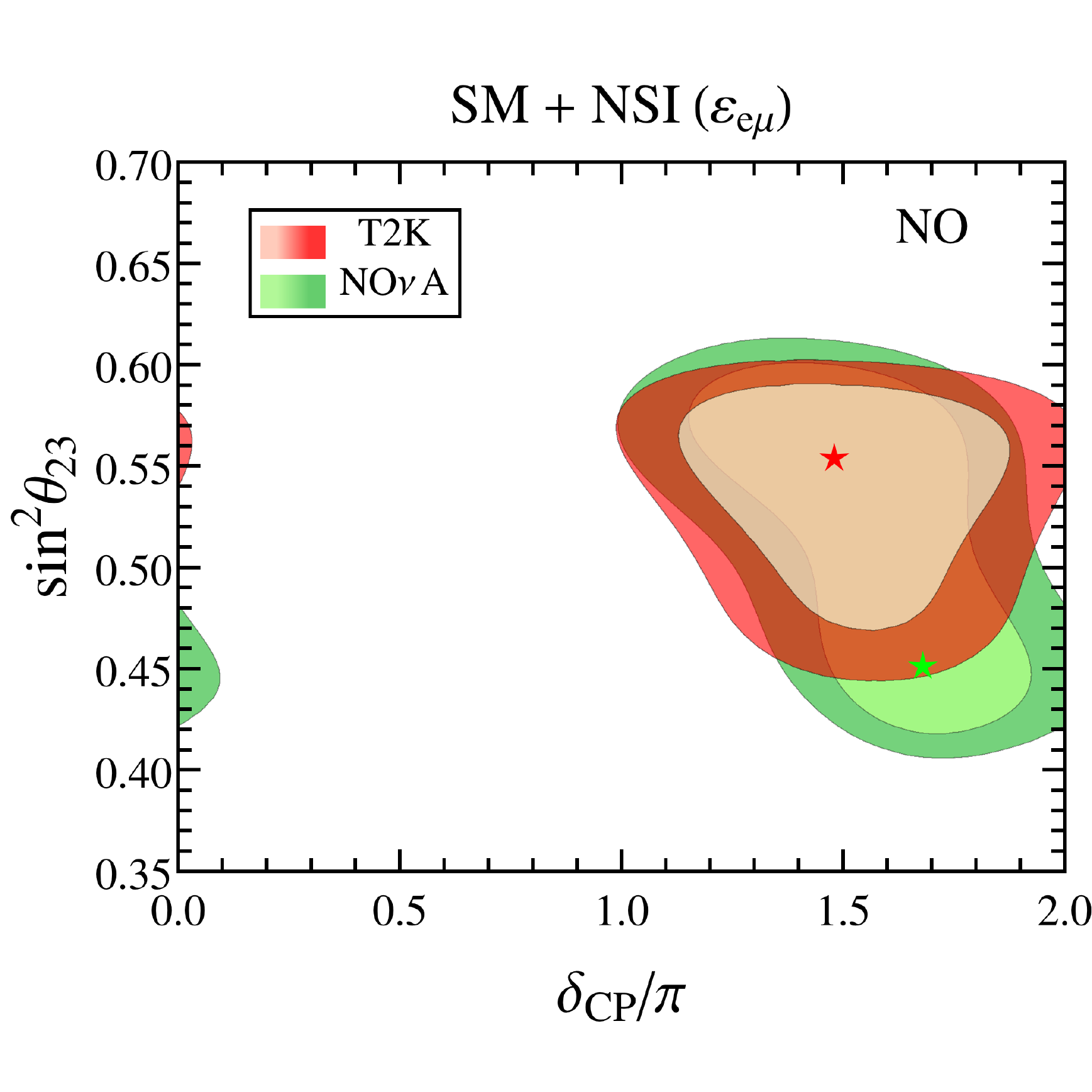}
\includegraphics[height=5.87cm,width=5.cm]{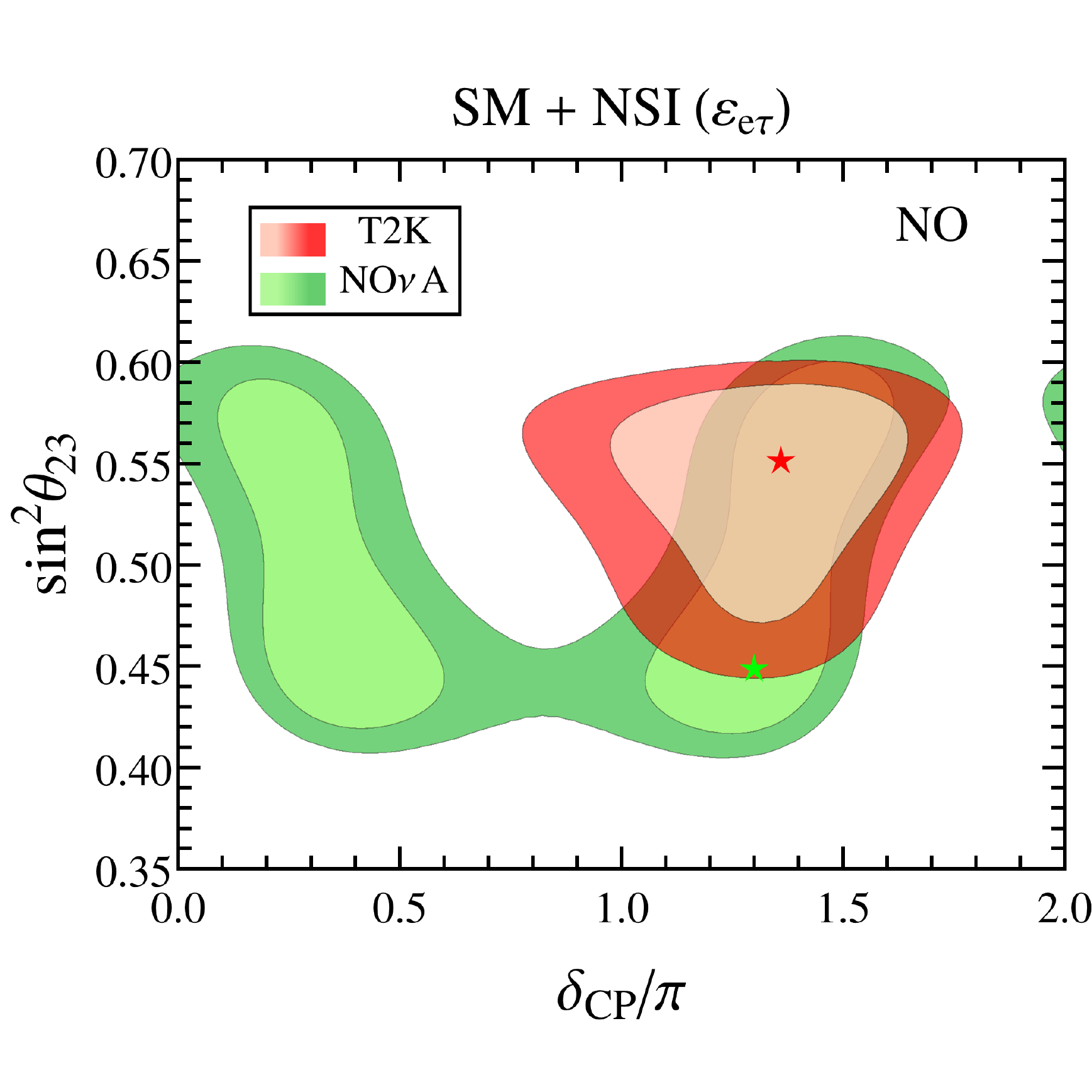}
\vspace*{-0.7cm}
\caption{Allowed regions derived separately by T2K and NO$\nu$A for NO in the plane $\left(\delta_{\rm CP}/\pi,\,\sin^2\theta_{23}\right)$. Left panel represents the SM case and middle (right) panel corresponds to the NSI in the $e-\mu$ ($e-\tau$) sector. In the middle panel we have considered 
the NSI parameters at their best fit values obtained from T2K + NO$\nu$A ($|\varepsilon_{e\mu}| = 0.15, \phi_{e\mu} = 1.38\pi$).
Similarly, in the right panel we have taken $|\varepsilon_{e\tau}| = 0.275, \phi_{e\tau} = 1.62\pi$.
The contours are drawn at the 68$\%$ and 90$\%$ C.L. for 2 d.o.f.. This figure has been taken from \cite{Chatterjee:2020kkm}.}
\label{fig_tension}
\end{figure*} 
%==================================================================

Let us now try to understand how the preference of the non-zero values of the NSI couplings in the $e-\mu$ and $e-\tau$ sectors help to resolve the tension between the NO$\nu$A and T2K measurement on $\delta_{\mathrm {CP}}$. In 
Fig.~\ref{fig_tension} we display the $68\%$ and $90\%$ C.L. allowed regions for 2 d.o.f. in the 
plane spanned by the standard CP-phase $\delta_{\mathrm {CP}}$ and the atmospheric 
mixing angle $\theta_{23}$ in the NO case. The left panel refers to the SM case,
while the middle and right panels concern the SM+NSI scenario with NSI in the $e-\mu$ and $e-\tau$ sectors
respectively. The red (green) contours correspond to T2K (NO$\nu$A) respectively. From the left panel it is clearly evident that the discrepancy between the NO$\nu$A and T2K lies at more than $90\%$ confidence level. NO$\nu$A (T2K) prefers values close to $\delta_{\mathrm {CP}} \sim 0.8 \pi$ ($\sim 1.4 \pi$). In the middle and right panel the contours have been drawn considering the best fit values of the NSI parameters obtained from the combined analysis of both the experiments as shown in Fig.~\ref{fig:regions_1} and Fig.~\ref{fig:regions_2} respectively. We can see that in presence of NSI both in the middle and right panel, the tension between the two experiments is resolved. The best fit values of $\delta_{\mathrm {CP}}$ are almost same for both the panel and they are close to $\sim 3\pi/2$. Interestingly, this is also the value very close to the one preferred by T2K. It can be understood from the fact that due to short baseline (295 km) T2K has low sensitivity to the matter effects and thereby to the NSI, whereas NO$\nu$A has a longer baseline (810 km) and hence more sensitive to matter effects as well as to NSI. As a result, the standard CP-phase $\delta_{\rm CP}$ measured by T2K can be considered more faithful. 
It is worth to mention that in the SM case, the IO is slightly preferred over NO ($\chi^2_{\rm NO}-\chi^2_{\rm IO}=1.87$). Now in presence of NSI, there is moderate preference of NO over IO in $e-\mu$ sector ($\chi^2_{\rm NO}-\chi^2_{\rm IO}=-2.56$), whereas in $e-\tau$ sector, there is no such preference ($\chi^2_{\rm NO}-\chi^2_{\rm IO}=-0.21$). For more details see~\cite{Chatterjee:2020kkm}.

{\bf {\em Conclusions.}} In this work, we explored the impact of NSI on resolving the discrepancy between the two long-baseline experiments T2K and NO$\nu$A in the measurement of standard CP-phase $\delta_{\rm CP}$. We found that this discrepancy can be resolved if one considers the flavor changing neutral current non-standard interactions of neutrinos of type involving $e-\mu$ and $e-\tau$ sectors. We hope that the future LBL accelerator data as well as the atmospheric data would shed more light in confirming the existence of NSI hypothesis.

{\bf {\em Acknowledgments.}} S.S.C. would like to thank the organizers of ``The 22nd International Workshop on Neutrinos from Accelerators (NuFact2021)'' for giving an opportunity to present this work.

\bibliographystyle{JHEP}
\bibliography{NSI-References_v2}

\end{document}